\documentclass[a4paper]{article}
\begin{document}
\title{Taking Kaluza seriously leads to a non-gauge-invariant electromagnetic theory in a curved spacetime}
\author{Luca Fabbri\\ Dipartimento di Fisica dell'Universit\`{a} di Bologna\\ Via Irnerio 46, 40126 Bologna, Italy}
\date{December 15, 2001}
\maketitle
\section*{abstract}
Kaluza's metric with the cylinder condition is considered without
the weak gravitational field approximation. It is shown that these
hypotheses lead to a non-gauge-invariant electromagnetic theory in
a curved space-time. The problem of electro-gravitational
unification is considered from this point of view.
\section{Introduction}
In the early Twenty's, Theodor Kaluza showed that by adding a
fifth dimension to space-time electromagnetism could be
geometrized side by side with gravitation, namely that provided
one assumed proportionality of the electromagnetic potentials to
the mixed components $(\mu 5)$ of the metric tensor, one could
derive the set of the equations of the Einstein-Maxwell
theory$^{(1)}$. Inasmuch as it gives nothing more that
electromagnetism \textit{and} gravitation, with no effective
extension of the physical picture, Kaluza's theory is subject to
the obvious remark that it does not produce anything new, thus
representing a purely formal transcription of already existing
theories.

Kaluza's results were however obtained in the weak field
approximation. A few years later Oskar Klein re-discovered the
fifth dimension, and, with an appropriate choice of the metric,
was able to re-obtain Kaluza's results$^{(2)}$. In this paper we
take the attitude to stick to Kaluza's original choice for the
metric, but to avoid taking the weak field approximation from the
start. Calculations shall be carried out exactly throughout, and a
weak field approximation shall be considered, as a final step,
only for the electromagnetic field.

Thus rephrased, Kaluza's framework proves richer than commonly
believed. It turns out that it does not merely reproduce the
Einstein-Maxwell theory, but describes a theory of the
electromagnetic field in an actually curved space-time, with an
extra electro-gravitational coupling. A notable feature is a
formal breaking of the electromagnetic gauge invariance, not
unlike the one taking place, for instance, in the transition to
superconductivity, exhibited by the form of the coupling and the
appearance of a mass term in the wave equation. I thought it
worthwhile to present this result not as much for its direct
physical interest but for its formal resemblance to other cases of
symmetry breaking.

\section{Hypotheses and assumptions}
Following Kaluza's basic idea we deal with a 5-dimensional world
to which we attribute a 5-dimensional Riemann geometry. We also
assume that the components of every tensor are independent from
$x^5$, that is $\partial_5 = 0$.

From now on, all our tensors will be projected on the
4-dimensional space-time, and we will use the convention that
every tensor with a subscript 5 on its left lives on the
5-dimensional space, while the same tensor with subscript 4 lives
on the 4-dimensional space. Latin capital letter indexes run from
1 to 5, Greek indexes from 1 to 4.

With these hypotheses we write the 5-dimensional metric as
\begin{equation}
g_{AB} = \left(
\begin{array}{ccc|c}
 & & & \\
 & g_{\alpha \beta} & & 2aA_{\alpha} \\
 & & &  \\ \hline
 & 2aA_{\beta} & & 1
\end{array}
\right)
\end{equation}
with $a$  a dimensionless constant.

We define the following two quantities
\begin{eqnarray}
F_{\alpha \beta} & = & \partial_{\alpha}A_{\beta} - \partial_{\beta}A_{\alpha} \\
\Sigma_{\alpha \beta} & = & \partial_{\alpha}A_{\beta} + \partial_{\beta}A_{\alpha}
\end{eqnarray}
where, according to Kaluza's interpretation, $F_{\alpha \beta}$ is
the electromagnetic field.
\\ \\
For the mathematical form of the 5-source tensor we generalize the
4-Energy-Momentum tensor of the matter, so we can write
\begin{equation}
T_{AB} = \sigma_{0} u_{A} u_{B}
\end{equation}
where $\sigma_{0}$ is the proper inertial mass density.

We may now define the following two quantities : \\
1) $\mu_0 \equiv \sigma_0$, where $\mu_0$ is the proper gravitational charge density; \\
2) $\frac{ac}{2} \rho_0 \equiv \sigma_0 u_5$, where $\rho_{0}$ is the proper electric charge density. \\

We observe that
\begin{itemize}
    \item[i)] since $\sigma_0$ is not negative and $u_5$ mag assume every value  we obtain that
    the gravitational charge cannot
    be negative, while the electric one can be positive, negative, or null;
    \item[ii)] writing $\rho_0 = \frac{2}{ac} \mu_0 u_5$, we can see massless charged particles cannot
    exist.
\end{itemize}

Projected on the 4-dimensional space-time and using these
definitions, the 1-source tensor is
\begin{eqnarray}
_5T_{\alpha \beta} & = & \mu_0 \: _5 u_{\alpha} \: _5 u_{\beta} \\
_5T_{5 \beta} & = & \frac{ac}{2} \: _5 u_{\beta} \: \rho_0
\end{eqnarray}
\\
Setting $G = 1$, we finally write the Einstein 5-dimensional
equations
\begin{equation}
G_{AB} + \frac{8 \pi}{c^2} T_{AB} = 0
\end{equation}
where $A$ and $B$ cannot be both equal to 5. These, projected on
the 4-dimensional space-time, split in a set of tensor equations
of order 2, and a set of vector equations
\begin{eqnarray}
_5G_{\alpha \beta} + \frac{8 \pi}{c^2} \mu_0 \: _5 u_{\alpha} \: _5 u_{\beta} & = & 0 \\
_5G_{5 \beta} + a \frac{4 \pi}{c} \: _5 u_{\beta} \rho_0 & = & 0
\end{eqnarray}

\section{Calculus and approximations}
Now we start the calculus of the Levi-Civita connection and of the
Riemann, Ricci and scalar curvatures; from the last two we obtain
the Einstein tensor.

For the Levi-Civita connection we find
\begin{eqnarray}
_5\Gamma_{\mu , \alpha \beta} & = & _4 \Gamma_{\mu , \alpha \beta} \\
_5\Gamma_{\mu , \alpha 5} & = & a F_{\alpha \mu} \\
_5\Gamma_{5 , \alpha \beta} & = & a \Sigma_{\alpha \beta}
\end{eqnarray}

For the Einstein tensor we obtain
\begin{eqnarray}
_5G_{\alpha \beta} & = & _4G_{\alpha \beta} + O_{\alpha \beta}(a^2) + O'_{\alpha \beta}(a^4) \\
_5G_{5 \beta} & = & a( _4 \nabla_{\mu} F_{\beta}\,^{\mu} -
A_{\beta} \: _4R ) + \Omega_{\beta}(a^3) +
    \Omega'_{\beta}(a^1)
\end{eqnarray}
where $O_{\alpha \beta}(a^2)$ and $O'_{\alpha \beta}(a^4)$ are two
tensors that behave as $a^2$ and $a^4$, when $a \longrightarrow 
0$; $\Omega_{\beta}(a^3)$ and $\Omega'_{\beta}(a^5)$ are two
vectors that behave as $a^3$ and $a^5$ in the same limit.
\\
We then make the approximation according to which the
dimensionless constant $a$ is very small, with the physical
interpretation that the electromagnetic field is not strong enough
as to contribute to the space-time curvature.

In this limit, we obtain
\begin{equation}
_5u_{\mu} = _4u_{\mu}
\end{equation}
so that
\begin{eqnarray}
_5T_{\alpha \beta} & = & _4T_{\alpha \beta} \\
_5T_{5 \beta} & = & \frac{ac}{2} \: _5u_{\beta} \rho_0 = \frac{ac}{2} \: _4u_{\beta} \rho_0
    = \frac{ac}{2} j_{\beta}
\end{eqnarray}
and
\begin{eqnarray}
_5G_{\alpha \beta} & = & _4G_{\alpha \beta}  \\
_5G_{5 \beta} & = & a( _4 \nabla_{\mu} F_{\beta}\,^{\mu} -
A_{\beta} \: _4R )
\end{eqnarray}
In this approximation, the Einstein's equations become
\begin{eqnarray}
_4G_{\alpha \beta} + \frac{8 \pi}{c^2} \: _4T_{\alpha \beta} & = & 0 \\
a ( _4 \nabla_{\mu} F_{\beta}\,^{\mu} - A_{\beta} \: _4R + \frac{4 \pi}{c} j_{\beta}) & = & 0
\end{eqnarray}
Omitting the subscript 4, we finally have Einstein's equations for
the gravitational field
\begin{equation}
G^{\alpha \beta} = - \frac{8 \pi}{c^2} T^{\alpha \beta}
\end{equation}
separated from the electromagnetic one as our approximation
requires, and a new kind of 4-vector equations
\begin{equation}
\nabla_{\mu} F^{\mu \beta} + A^{\beta} R  = \frac{4 \pi}{c} j^{\beta}
\end{equation}
We observe that :
\begin{itemize}
    \item[i)] In the electromagnetic vacuum one has
        \begin{equation}
        \nabla_{\mu} F^{\mu \beta} + A^{\beta} R = 0
        \end{equation}
    linear in the electromagnetic potential.
    \item[ii)] taking the covariant quadridivergence of this equation
    gives a subsidiary condition of the form
        \begin{equation}
        \nabla_{\mu} ( A^{\mu} R  ) = 0
        \end{equation}
    Indeed, from the identity
        \begin{equation}
        \nabla_{\mu} \nabla_{\lambda}S_{\alpha \beta} -\nabla_{\lambda} \nabla_{\mu}S_{\alpha \beta}
            \equiv S_{\alpha \nu} R^{\nu}\, _{\beta \lambda \mu} +
             S_{\nu \beta} R^{\nu}\, _{\alpha \lambda \mu}
        \end{equation}
    where $R^{\nu}\, _{\beta \lambda \mu}$ is the Riemann curvature and $S_{\alpha \beta}$ a
    generic tensor, one obtains with a little algebra
        \begin{equation}
        \nabla_{\mu} \nabla_{\alpha}S^{\mu \alpha} \equiv
            \nabla_{\alpha} \nabla_{\mu} S^{\mu \alpha}
        \end{equation}
    If $S_{\alpha \beta}$ is an antisymmetric tensor one then has
        \begin{equation}
        \nabla_{\mu} \nabla_{\alpha}S^{\mu \alpha} \equiv 0
        \end{equation}
     This identity and the continuity equation for the electric
     current $\nabla_{\alpha} j^{\alpha} = 0$ lead to the
     subsidiary condition.

    With this subsidiary condition, the field equations read
        \begin{eqnarray}
        \nabla_{\mu} F^{\mu \beta} + A^{\beta} R & = & 0 \\
        \nabla_{\mu} ( A^{\mu} R  ) & = & 0
        \end{eqnarray}
    and describe a non-gauge-invariant 4-vector massive field.
\end{itemize}
In the gravitational vacuum we have $R \equiv 0$ and the equations
for the electromagnetic field reduce to
\begin{equation}
\nabla_{\mu} F^{\mu \alpha} = \frac{4 \pi}{c} j^{\alpha}
\end{equation}
Furthermore, we have the identities
\begin{equation}
\epsilon^{\alpha \beta \gamma \delta}  \nabla_{\gamma} F_{\alpha \beta} = 0
\end{equation}
i.e. the whole set of Maxwell's inhomogeneous and homogeneous
equations in a curved space-time.

\section{Summary}
As we have already observed, we have a set of 5 equations for the
electromagnetic field in the matter
\begin{eqnarray}
\nabla_{\mu} F^{\mu \alpha} + A^{\alpha} R & = & \frac{4 \pi}{c} j^{\alpha} \\
\nabla_{\mu} (A^{\mu} R) & = & 0
\end{eqnarray}
This a Proca-like system of equations  describing a non-gauge-
invariant 4-vector massive field. This can be interpreted in
this way: when an electromagnetic field propagates in a space-time
with a non-vanishing scalar curvature, it couples with the
gravitational one and this coupling brakes the gauge symmetry,
giving a mass to the photon.

However, since $R$, calculated in a fixed point, is different from
$0$ if and only if the proper inertial mass density does not
vanish in that point, and since when $R \neq 0$ the matter
undergoes a gravitational collapse, one has a hint as to why, even
assuming the validity of the present framework, the
non-gauge-invariant term does not show up in  Maxwell's equations
and why the photon mass has always been found equal to $0$.

The non-gauge-invariance of this theory is also interesting for a
discussion about the problem of unification. Following Pauli,
every theory which is generally covariant and gauge invariant can
be formulated in Kaluza's form; although this theory is generally
covariant, it is not gauge invariant, and thus, it cannot be
formulated in Kaluza's form unless this possibility is warranted
by its own structure. \\

For a more general discussion about unified theories we can recall
Lichnerowicz's definition of a unified theory$^{(3)}$:

\textit{ "A theory is unified in a broad sense if, in the
representation of the fields and in the form of the equations, it
attributes symmetrical roles to the two fields; in particular, in
the conceptions of general relativity, the two fields emanate from
the same geometry. A theory is unified in a strict sense if the
exact equations govern a non-decomposable hyperfield, and they can
only approximately be decomposed into two field equations when one
of the fields dominates the other."}

From this point of view, both Kaluza's theory and this extension
satisfy Lichnerowicz's definition of a unified theory in the broad
sense, since the gravitational and the electromagnetic field
emanate from the same geometry. However, Kaluza's original theory
is nothing more than the Einstein-Maxwell theory, while, in this
extension, we have Einstein and Maxwell theories in a curved
space-time as the approximated form of a 4-tensor theory and of a
4-vector one which are two different space-time projections of a
same 5-tensor theory.

From these considerations we obtain, as Pauli already
said$^{(4)}$, that the Kaluza's theory is not a unified theory in
the strict sense; but from the same considerations we also obtain
that this theory is a unified theory in the strict sense.

\section*{Acknowledgments}
I would like to thank prof. S.Bergia for encouragement and help,
dott.A.P.Franco, R.Govoni and F. Nironi for critical comments and
computer support.

\section*{References}
\begin{itemize}
\item[1.] Th. Kaluza, \textit{Zum Unit$\ddot{a}$tsproblem der Physik,
   Sitzungsber. D. Berl. Akad.} 465 (1918). English translation in
   T. Appelquist, A. Chodos and P.G.O. Freund, \textit{Modern
   Kaluza-Klein Theories} (Addison-Wesley, 1987) pp.61-68.
\item[2.] O. Klein, \textit{Quantentheorie und
F$\ddot{u}$nfdimensionale
   Relativit$\ddot{a}$tstheorie, Zeitschrift f$\ddot{u}$r Physik}
   \textbf{37} 895-906. English translation in
   T. Appelquist, A. Chodos and P.G.O. Freund, \textit{Modern
   Kaluza-Klein Theories} (Addison-Wesley, 1987) pp.76-87.
\item[3.] A. Lichnerowicz, \textit{Theories Relativistes de la
   Gravitation et de l'Electromagnetisme} (Masson, 1955).
\item[4.] W. Pauli, \textit{Theory of Relativity} (1958).
\end{itemize}
\end{document}